\documentclass[twocolumn]{revtex4}
\usepackage{graphicx}
\usepackage{color}

\begin{document}

\title{Artifacts of opinion dynamics at one dimension }
\author{Serge Galam}
\affiliation{Centre de Recherche en \'Epist\'emologie Appliqu\'ee,\\
\'Ecole Polytechnique and CNRS, \\CREA, Boulevard Victor, 32,
75015 Paris, France}
\altaffiliation{serge.galam@polytechnique.edu}
\author{Andr\'e C. R. Martins}
\affiliation{GRIFE -- EACH -- Universidade de S\~ao Paulo,\\
Rua Arlindo Bétio, 1000, 03828--000,  S\~ao Paulo, Brazil}
\altaffiliation[Also at ]{Centre de Recherche en \'Epist\'emologie Appliqu\'ee,
\'Ecole Polytechnique, CREA, Boulevard Victor, 32,
75015 Paris, France}

\begin{abstract}

The dynamics of a one dimensional Ising spin system is investigated using three families of local update rules, the Galam majority rules, Glauber inflow influences and Sznadj outflow drives. Given an initial density $p$ of up spins the probability to reach a final state with all spins up is calculated exactly for each choice.  The various formulas are compared to a series of previous calculations obtained analytically using the Kirkwood approximation. They turn out to be identical. The apparent discrepancy with the Galam unifying frame is addressed.  The difference in the results seems to stem directly from the implementation of the local update rule used to perform the associated  numerical simulations. The findings lead to view the non stepwise exit probability as an artifact of the one dimensional finite size system with fixed spins. The suitability and the significance to perform numerical simulations  to model social behavior without solid constraints is discussed and the question of what it means to have a mean field result in this context is addressed.

\end{abstract}

%Key words: Sociophysics, opinion dynamics, Ising spins, one dimensional chain

\maketitle

In this Letter, we reinvestigate the dynamics of an extremely simple opinion dynamics model, a slightly modified version of the original Sznadj outflow rule \cite{sznajd} proposed by Slanina et al \cite{slanina1}, which was shown to exhibit a rich variety of behaviors. In particular, when calculating analytically  the probability to reach a uniform final state, they found it is a continuous function of the initial magnetization. The result was confirmed by Monte Carlo simulations. Starting from a generalized voter model an independent and simultaneous paper by Lambiotte and Redner \cite{redner1} recovered both results, analytically and numerically. The existence of a continuous exit probability contradicts the prediction of the Galam unifying frame (GUF), also denoted as the general sequential probabilistic frame (GSPF) \cite{unify}, which yields a threshold stepwise function for the exit probability.

We show that these analytical results are obtained at once by treating one single triplet instead of the total system. Alternatively, a straightforward one shot extended voter model is found to reproduce the same results. The two treatments are a direct application of classical mean field technics, which indeed are based on a ``one-site approach"  \cite{mean-field}.  In addition, a simple sequential update is found to be sufficient to recover the results via a Monte Carlo simulation. 

Based on the results of this mean-field approach applied to triplets, we arrive at three conclusions: (i) The Kirkwood approximation \cite{kir} treatment of the Slanina chain (and, by extension, and the Lambiotte and Redner treatment) reduces to a simple mean field approach. (ii) Contrary to several author claims, the GUF does go beyond the mean field. Indeed a one step GUF restricted to an average single interactive group yields the mean field level. (iii)  The Monte Carlo simulations, which were found in perfect agreement with the analytical formula, must have been biased either by finite size effects or by using an inappropriate sequential update.

To set the stage for the demonstration, we recall that, in statistical physics, microscopic interactions are given by defining the Hamiltonian of the system and the associated macroscopic properties obtained from analytical calculations or from simulations that are expected to be independent of the choice of the dynamics used to implement those interactions. Moreover, while performing simulations, a special attention is given in the actual choice of the update procedure,  either sequential or simultaneous or random, to avoid results which could be artifacts of the procedure used, with the system being trapped in some specific configurations.

However, in opinion dynamics, most models focus on the choice of the update rule without formulating the actual interactions \cite{fortunato}. Some exceptions exist \cite{strike, mosco}. It is similar in economy with game theory where only the outcomes are given without an explicit formulation of the interactions \cite{walliser}. The social content of the model is exhibited in the setting of the update rule, which is supposed to incorporate the complex and unknown mechanisms of the underlying cognitive interactions. Understanding what a mean-field approach is under these circumstances can be difficult. This is probably the cause for the confusion about above approaches, which turn out to be approximations.

The fact is that a great deal of models have been proposed to describe opinion forming, each one displaying a specific update rule \cite{sznajd, bounded, martins1, martins2, sousa, ausloos, pierluigi}. Usually, this makes it difficult to perform analytic calculations and most of the results are obtained from numerical simulations where the rule can be implemented. The significance of the model is then discussed on the sole basis of the numerical results, often obtained from small size samples, usually of the order of 100 or 1000, rarely more. Such a practice, combined with the almost impossible making of real experiments, renders it difficult to assess the social relevance of a model without solid ground. Even the validity of the numerical results cannot be certain since they are not constrained by some general accepted principles, which have to be obeyed like in physics. Another difference is that in statistical physics interactions are simultaneously activated among the spins although they usually involve only nearest neighbors while in social systems interactions are restricted to separate groups of agents at a time via local interactions among the agents. 

Here, we first outline the basic features of the  modified Szandj model with the approximations used to solve it. The corresponding results obtained by analytical calculations and numerical simulations are listed. Then we present Galam exact derivation of the same results and seek to understand three basic issues: (i) What is the physical implications of that identity? (ii) What is the social significance of the associated Monte Carlo simulations. (iii) What is the validity of an exit probability which is not stepwise? We demonstrate that, while GUF provides a mean field result if iterated only once, it is not mean field when iterated more than once and to several local groups. Applied to the Slanina model the one step averaged GUF reproduces the same result reported previously \cite{slanina1, redner1}. 

\textit{The Slanina et al model and results.} Consider $N$ agents on a linear chain at sites denoted respectively $1, 2, ..., N$. A two state variable $S_i=\pm 1$ with $i=1, 2, ..., N$ is attached to the agent at site $i$, which represents its current opinion, either $+1$ or $-1$. Given a distribution of individual opinions the system evolves by repeated applications of the following procedure: (a) Two neighboring agents are selected randomly. (b) In case they have different opinions, nothing happens and a new pair is selected. (c) If they do share the same opinion, one neighbor of the pair is chosen randomly with equiprobability. (c) The selected neighbor adopts the common opinion, if it didn't already agree to that. (d) The process is repeated till all agents hold the same opinion. While two opinions can flip simultaneously in the original Sznadj formulation, here one spin is flipped at most in one step. 

Given a distribution of $+1$ and $-1$, the probability $P_+$ to have all agents sharing the opinion $+1$ is calculated as a function of the density $p_0$ of $+1$, which are initially present in the chain. It is called the exit probability. The number of updates required to reach the state with all agents aligned is also calculated.

The calculation of the exit probability is carried on applying a decoupling approximation to truncate the equation hierarchy for multi-point spin correlations. It is used in various contexts with different names.  The one considered was introduced by Kirkwood to study the statistical mechanics of fluid mixtures \cite{kir}. It is similar to the Hartree-Fock approximation but is more limited since it cannot  be systematically improved with diagrammatic techniques. A second approximation is applied with the assumption of  low decay of correlations. Nearest neighbor correlations are extended up to next next nearest neighbors. After elaborated manipulations the following exit probability is obtained with
\begin{equation}\label{eq:slanina1}
P_{+}\approx \frac{p_{0}^2}{2p_{0}^2- 2p_{0}+1}  .
\label{s1}
\end{equation}

\begin{figure}
\includegraphics[width=.45\textwidth]{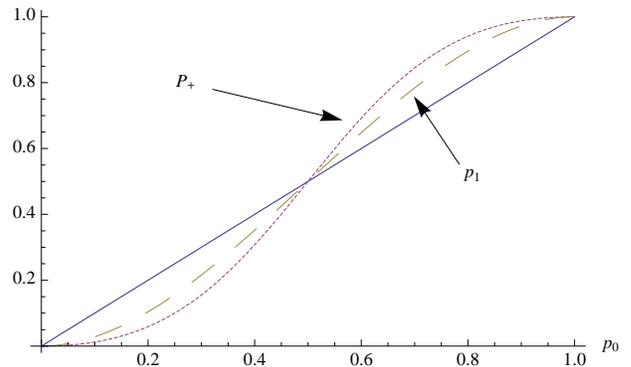}
\caption{The variations of both $P_{+}$ from Eq. (\ref{eq:slanina1}) and $p_1$ from Eq. (\ref{}) are exhibited as a function of $p_{0}$.}
\label{fig:slanina1}
\end{figure}

The function is shown in Figure (\ref{fig:slanina1}) and was found in perfect agreement with Monte Carlo simulations of chains of sizes up to $10^3$ agents with $10^4$ sampling \cite{slanina1}. Identical results were obtained in \cite{redner1} with 100 agents averaged over $5\times 10^3$ realizations.

\textit{The novel treatment.} In order to get a mean field result to the model, we first reformulate it by treating a triplet instead of a pair plus a random neighbor. Pick at random one site $S_i$. Then complete the triplet with equal probability, either on its left with $S_{i-2}, S_{i-1}$ or on its right with $S_{i+1}, S_{i+2}$. In the first case, if $S_{i-2}=S_{i-1}=a$, then $S_i=a$ and in the second case if $S_{i+1}=S_{i+2}=b$, then $S_i=b$ with $a, b=$ + or -. Otherwise for $S_{i-2}\neq S_{i-1}$ or $S_{i+1}\neq S_{i+2}$, nothing happens and a new site is selected. We consider that one interaction has happened only when one site does change its value.

Accordingly, the probability to have a selected site $i$ to end +, is just $\frac{1}{2}p_{0}^2+\frac{1}{2}p_{0}^2=p_{0}^2$, the probability to have a pair of + on the right or on the left. By symmetry the probability to have a selected site $i$ to end -, is just $(1-p_{0})^2$, the probability to have a pair of -. However, since the case of a mixed pair $S_{i-2}=- S_{i-1}$ or $S_{i+1}=-S_{i+2}$ is dismissed with a new site $S_i$ being selected, the sum of above two probabilities is not equal to one. Therefore they must be rescaled by their sum to get normalized to one. Given above update rule the normalized probability to have a site updated to + is thus,
\begin{equation}
p_{1}= \frac{p_{0}^2}{p_{0}^2+(1- p_{0})^2} .
\label{g1}
\end{equation}

The same result is obtained by picking randomly a pair of adjacent sites. If they are in different states, select another pair. The probability of this event is $2p(1-p)$. Otherwise, if they share the same state, i.e. + or - with respective probabilities $p^2$ and $(1-p)^2$, select the left or right adjacent site and align it along the pair common state. The normalized probability to get an updated + site is thus given also by Eq. (\ref{g1}).

At this stage, it is worth stressing that Eq. (\ref{g1}) is identical to Eq. (\ref{s1}). However while $P_{+}$ is the probability to end up with all sites at + after a series of repeated updates, $p_{1}$ is the probability to get one site + after one update for one selected site and triplet. The next step is then to calculate the exit probability which results from successive iterations of Eq. (\ref{g1}) for all sites. 

One possible way to implement the calculation is by using the following sequential update procedure: Pick up one pair of adjacent sites, if they are heterogeneous, pick another pair. Once the selected pair is homogenous, align a random adjacent site along the common state, either on the left or on the right.  To illustrate the procedure, consider the case where the pair is ++ with
\begin{equation}
... \bullet   \bullet  \bullet (++) \bullet \bullet  \bullet \bullet   ...  ;
\label{pair-k} 
\end{equation}
Select the adjacent site to be updated on the right (upper line) or on the left (lower line),
\begin{equation}
... \bullet   \bullet  \bullet (++) \bullet \bullet  \bullet \bullet   ... \rightarrow
\left\{
\begin{array}{ccc}
... \bullet   \bullet  \bullet (++ \bullet) \bullet  \bullet \bullet   ...   \\ \\
... \bullet   \bullet  (\bullet ++) \bullet \bullet  \bullet \bullet   ... 
\end{array} 
\right. ;
\label{pair-k} 
\end{equation}
Update for instance the right upper line). It yields, 
\begin{equation}
 ... \bullet   \bullet  \bullet (+++) \bullet \bullet \bullet   ...  \  ;
\label{g2}
\end{equation}
Move the triplet one site left or right as follows,
\begin{equation}
... \bullet   \bullet  \bullet (+++) \bullet  \bullet \bullet   ... \rightarrow
\left\{
\begin{array}{ccc}
... \bullet   \bullet  \bullet +(++ \bullet)  \bullet \bullet   ...  \\ \\
... \bullet   \bullet  (\bullet ++)+ \bullet  \bullet \bullet   ... 
\end{array} 
\right. ;
\label{pair-k} 
\end{equation}
Update one of the triplet with one of the two chains,
\begin{equation}
\left\{
\begin{array}{ccc}
... \bullet   \bullet  \bullet +(+++)  \bullet \bullet   ...  \\ \\
... \bullet   \bullet  (+ ++)+ \bullet  \bullet \bullet   ... 
\end{array} 
\right. ;
\label{pair-k} 
\end{equation}

Repeat the process by shifting one site left or right the triplet as above. The whole chain will end up with all sites in state +. Accordingly, the only random step is the selection of the initial pair, then the process is deterministic with respect to the final alignment. It proves that the sequential update procedure leads to $P_{+}=p_1$.

The elaborated Kirkwood decoupling approximation to truncate the equation hierarchy for multi-point spin correlations is thus identical to above straightforwards calculation. 

\textit{The one step mean field voter model extension.}  A similar analysis can be performed by using the voter model as basis. Voter model is based on a conservative dynamics using a pair directed interaction \cite{voter}. Select one site $S_i$ at random, pick a second one $S_j$ again at random, then update the first site putting $S_i=S_j$ with $S_j$ unchanged. Within the GUF, the voter model was shown to be the frontier between the ordered phase, characterized by the coexistence of a majority and a minority, including the extreme limit of a zero minority, and the disordered phase where a perfect equilibrium between the two opinions prevails \cite{unify}. 

A mean field treatment of the voter model \cite{mean-field} consists of treating one site exactly while the interacting sites are taken at their mean value. As the system is at zero temperature, there is no thermal variations in the sites that are treated exactly and we need to treat the problem differently. With no probabilistic distribution of spins due to thermal noise, the only probabilistic aspect left is that of random distribution of spins on the sites. The only two possible equilibria are when the pair site-external field (external site) have the same spin. As both have the same probability of being up, initially, the chain is all + with the probability $p_0^2$ and all - with probability $(1-p_0)^2$. However since the case of a mixed pair is dismissed, both probabilities should be normalized, which reproduces $P_+=p_1$ given by Eqs. ({\ref{s1}, \ref{g1}). This restriction to one step update of one site produces a mean field result. Our demonstration enlightens the mean field nature of the previous derivation. 

The only difference between the various calculations is the number of Monte Carlo steps needed to reach the final sate with all sites sharing the same opinion. These numbers are a direct function of the procedure chosen to implement the updates. They are not robust but could be used to make social predictions with respect to setting some procedural frame for decision making. Indeed, it is an efficient way to gain a lot of time to reach the same final decision. 

\begin{figure}
\includegraphics[width=.45\textwidth]{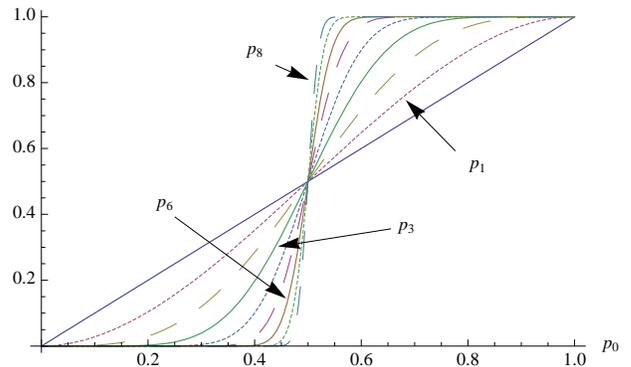}
\caption{The successive iterations of Eq. (\ref{}) with $p_{1}, p_{2},p_{3}, ...p_{8}$ as a function of $p_{t}$. The drive towards a step function for the exit probability is clearly seen.}
\label{k=1/2}
\end{figure}

\textit{Applying the GUF.} Above results help to understand the discrepancy noticed by Slanina about the GUF, which is expected to yield a step function for $P_+$, while other approaches have provided a continuous function. We have seen that the results of those approaches can be obtained from the analysis of a very small number of interacting sites, by using the GUF approach and can be seen, in each case, as a consequence of the method used to obtain the result, instead of the correct thermodynamic solution. 

In the case of the modified Sznajd rules, the Kirkwood approximation produced the same result for $P_+$ as the probability $p_1$ that the first spin to change in the system will change to $+$. And we have shown that in a simulation, depending on the update procedure, that probability can be the actual final probability that the whole system will end as $+$, but that this was obtained as an artifact of the implementation rule and not a correct result. Another possibility is that the observed simulation results could be a consequence of finite size problems. Indeed, the exact result for $N=3$ sites is very easy to obtain as the system is completelly deterministic. It is basically the probability that the randomly drawn spins will have a majority of $+$s and we get $P_+=3 p_0^2 - 2 p_0^3$. This curve is further away from the step function than Kirkwood approximation. As the $N$ becomes larger, $P_+$ should get closer to the Kirkwood solution in its path to a step function and this might be another cause for the observed behavior of the simulations.

In summary, we have demonstrated that Slanina and Lambiotte results can be obtained from a mean-field approach. On the other hand, the GUF treatment of the same problem does provide a continuous function, qualitatively similar to the ones obtained by the Kirkwood approximation, if one uses only one interaction. However, the sites will keep updating their spins and, after a few iterations, it is easy to see that the GUF approach tends to a step exit probability.  The additional subsequent steps drive the system further apart from mean field-like result and this shows that GUF is actually not a mean field result. This helps understand better the subtleties of extending the concept of a mean field result to a zero temperature problem. 

We conclude that when the  hamiltonian of the system is not properly defined and, instead, update rules are provided, special care is needed in order to avoid unwanted artifacts of the implementation details. Finite size effects with the system still out of equilibrium might be far more important than simulations might indicate. It is true that social systems are not so large that finite size systems will be completelly avoided and, in that sense, the previous results of Slanina and Lambiotte can have significative consequences in real systems. But, from a Physics perspective, they are probably consequences of updates rules and the finite size effects.

\textit{Acknowledgements.} One of the authors (ACRM) would like to thank the Funda\c{c}\~ao de Amparo a Pesquisa do Estado de S\~ao Paulo (FAPESP) for the support to the work under grant 2009/08186-0.

%%%%%%%%%%%%%%%%%%%%%
%To be completed

\end{document}